# A Comparative Study of Hidden Web Crawlers

Sonali Gupta[#1], Komal Kumar Bhatia

[#]*Associate Professors & Department of Computer Engineering, YMCA University of Science & Technology*
*Faridabad, India*

*Abstract*— **A large amount of data on the WWW remains inaccessible to crawlers of Web search engines because it can only be exposed on demand as users fill out and submit forms. The Hidden web refers to the collection of Web data which can be accessed by the crawler only through an interaction with the Web-based search form and not simply by traversing hyperlinks. Research on Hidden Web has emerged almost a decade ago with the main line being exploring ways to access the content in online databases that are usually hidden behind search forms. The efforts in the area mainly focus on designing hidden Web crawlers that focus on learning forms and filling them with meaningful values. The paper gives an insight into the various Hidden Web Crawlers developed for the purpose giving a mention to the advantages and shortcoming of the techniques employed in each.**

## I. INTRODUCTION

Majority of the Internet users depend on the use of search engines like Google, Yahoo, and Bing etc. to find the information on the Web. Most of these search engines provide entry only to the Surface Web, which is a part of the Web that can be discovered by following hyperlinks and downloading the snapshots of pages for including them in the search engine's index [1]. The results provided by the search engine are based in this copy of its local index.

Perhaps an even larger amount of information is available in the Hidden Web, which is a part of the WWW that cannot be discovered by simply following the hyperlinks. A simple example of this content includes the structured databases like product or online library catalogs, satellite images that are offered by search websites which can be accessed by submitting a search form. Another category of hidden Web content includes the dynamic data provided by web applications which give real-time information based on a particular user request like the online travel planners or booking systems. The same request when issued at different times result in different information. Although, these websites may provide a hyperlink structure to the database items so as to accommodate crawling by the crawlers designed for the surface web. But this does not guarantee that those search engines will have the current and updated information on prices and items in stock. Intuitively, this significant portion of the Web containing publicly available information in the form of electronic web databases is poorly accessible by conventional crawlers designed for general purpose search engines. Thus, in the literature we have a relevant class of crawlers that effectively work on retrieving and accessing this hidden information in databases, termed the Hidden Web Crawlers [3], [6], [12].

## II. CHARACTERISTICS AND SCALE OF THE HIDDEN WEB

The Hidden Web provides access to huge and rapidly growing data repositories on the Web. Some authors have obtained approximations to its huge size: In 2001, an initial study by Bergman indicated the size of the data in the Hidden Web to be approximately 500 times the size of the data in the Surface Web which included as many as 43,000-96,000 web sites offering access to 7500 terabytes of data [1]. Later in 2004, Chang et.al. Used a random IP sampling approach to measure the Hidden web content in online databases and revealed that most of the data in such databases is structured [15]. Further in 2007, Ben He et.al. by analyzing the percentage overlap between the most commonly used search engines such as Yahoo!, Google and MSN discovered the number of such sites to 236,000- 377,000 with only 37% of the available content being indexed by these search engines [16]. Thus, according to experts, the hidden Web forms the largest growing category of new information on the Internet and comprises of:

➢ Nearly 550 billion documents
➢ Content high relevant to every information need, market and domain
➢ Up to 2,000 time's greater content than that of the Surface Web.
➢ 95% publicly accessible information not subject to fees or subscription.
➢ More focused content than Surface Web sites.

## III. ACCESSING THE HIDDEN WEB

A user accesses the data in the Hidden Web by issuing a query through the search form (an interface provided by the Web site), which in turn gives a list of links to relevant pages on the Web. The user then looks at the obtained list and follows the associated links to find interesting pages. These search forms have been designed primarily for human consumption but serve as the only entry point to the Hidden Web, thus must be modelled and processed. There are two basic approaches to this end:

➢ Surfacing refers to the crawler's activity of collecting in the background as much relevant and interesting fraction of the data as possible and updating the search engine's index. The Hidden Web crawler has to automatically process the search forms after downloading it form the hidden web site and submit the filled form so as to download the response pages which can then be used with existing index structures of the search engine. This approach has the main advantage of best fit with the conventional search engine technology. Though pre-computing the most relevant form submissions for all interesting HTML forms is a challenging issue but is a passive task that can be carried off-line by the crawler when active, independent of the run-time characteristics of the hidden web resources. Thus, the approach is straightforward and is easily applicable.

➢ Virtual Data Integration which refers to the creation of a specific virtual schema for each domain and mapping the fields of the search forms in that domain to the attributes of the virtual schema. This enables the user to query over all the resources in its domain of interest just by filling a single search form in the domain. Search systems using such vertical schema are called vertical search engines. APIs are then used to access Hidden Web sources at query time and construct the result pages based from the retrieved responses. As external API calls need to be made by the search engine , the process relies on the performance of the Hidden Web sources, involving access latency thereby making it slower than traditional crawling or Surfacing .

The biggest challenge here is creating & generating a mediated schema and the semantic mappings between individual data sources and the mediator form. The problem has been termed as query routing. In particular, the queries on any search engine typically is a set of keywords reformulating which requires identifying the relevant domain of a query and appropriately routing the keywords in the query to the fields of the virtual schema that has been designed for the candidate domain.

Moreover, the number of domains on the Web is very large and precisely defining boundaries for a domain is tricky making the design of virtual schemata even more challenging.

Although research has been done in the area of developing web integration systems but the technological difficulties involved in the integration approach guide us to choose the approach of Surfacing as the road to success and discussed hereafter.

## IV. APPROACHES FOR SURFACING THE HIDDEN WEB

The crawler to extract the content in the Hidden Web has to imitate the above described set of steps that are being followed by the human i.e. the crawler when provided with the search form has to generate a query, issue it to the Web site, download the result index page, and follow the links to download the actual pages. The authors in [12] have proposed the following generic algorithm for any Hidden Web crawler.

```
Algorithm Hidden_web_crawl()

Step1: While (Resources
available)
do
2. qi= SelectTerm()
//select a term to send to the
site
3. R(qi) = QueryWebSite( qi)
/*where qi is the selected
query & R(qi) is the result page
for the Query qi.*/
4. Download ( R(qi));
5. End;
```

Fig. 1  Algorithm for crawling Hidden Web Site.

Crawling the Hidden Web involves two prime tasks of resource discovery and content extraction. The former deals with automatically finding relevant Web sites that contain a search form interface while the latter deals with obtaining the information from these sites by filling out forms with relevant queries or keywords. Circumscribed by the crawler's limitation of resources and the huge size of the Hidden Web, the common approach to crawl in the contents of the Hidden Web involves:

1) *Breadth-Oriented crawling:* As the hidden Web contains tens of millions of databases and search forms, a breadth oriented hidden Web crawler focuses on covering more and more data sources rather than exhaustively crawling the content inside one specific data source. Thus, the major challenge in this kind of crawling seems to be locating the hidden Web resources and analyzing the returned results for learning and understanding the interface required to automate the process of content extraction.

2) *Depth-Oriented crawling:* It focuses on extracting the contents from a designated hidden web resource i.e. the goal is to acquire most of the data from the given data source. Now, the crucial challenge for the crawler is to actively issue queries at the search interface of the designated database in order to uncover the database contents while incurring minimal cost. However, the crawler must automatically generate promising queries so as to carry out efficient crawling which is an exigent task. The problem is termed as query selection.

Perhaps, the above approaches are equally facilitated by gaining an insight into the type of information being contained in any web database which may be categorized either as unstructured or structured. Unstructured databases usually contain plain-text documents which are not well structured and provide a simple keyword-based search interface having an input control (text type) where users type a list of keywords to fill it in. Fig. 2 shows an example of such a search interface.

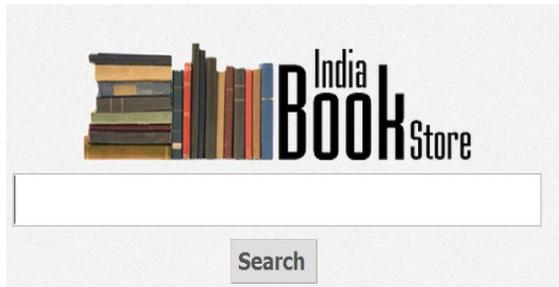

Fig. 2 Keyword-based Search Interface

In contrast, structured databases provide multi-attribute search interfaces that have multiple query boxes pertaining to different aspects of the content. For example, Fig. 3 shows a multi-attribute search form interface for an online book store offering structured content (title, author, publisher, price, ISBN, number of pages) coupled with a structured query interface (typically a subset of the content attributes like title, author, ISBN, publisher).

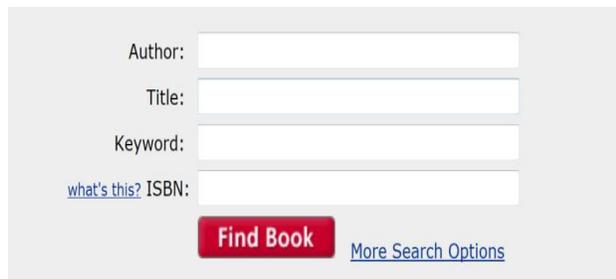

Fig. 3 A multi-attribute search form interface for an online book store

## V. Hidden Web Crawlers

Crawling techniques have been studied since the advent of the Web itself but the research on Hidden Web crawlers emerged with pioneering work by Raghavan and Molina in 2001. They have focused on a design for extracting content from electronic databases. Since, then numerous depth-oriented Hidden Web crawlers for structured as well as unstructured databases have been framed and developed, a review of which has been presented in the section.

### A. Depth-oriented Crawlers for structured databases

Raghavan in 2001 introduced the problem by presenting an operational model shown in Fig. 4 to describe the interaction that takes place between the crawler and the search form [3]. This model serves as a basis for their prototype hidden Web Crawler called the HiWE (Hidden Web Exposer), an outline of the architecture of which is given in Fig. 5. They have proposed a method for filling up search forms by raising potential queries that are either provided manually or collected from the query interfaces. The term form page is used to denote the page containing a search form and response page is used to denote the page received in response to a form submission.

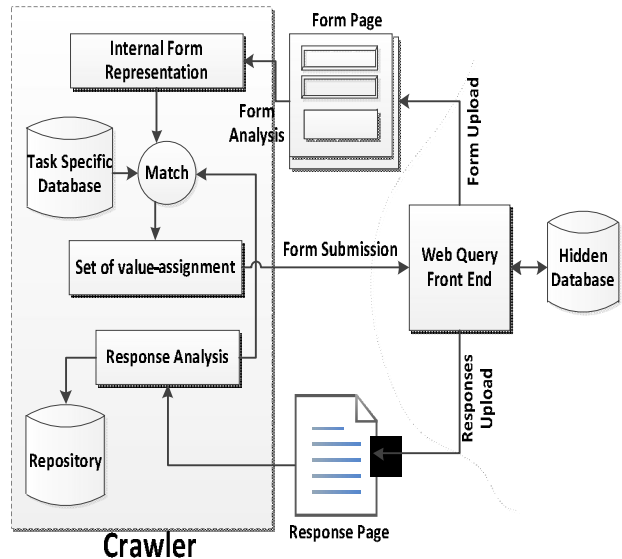

Fig. 4 Crawler Form Interaction

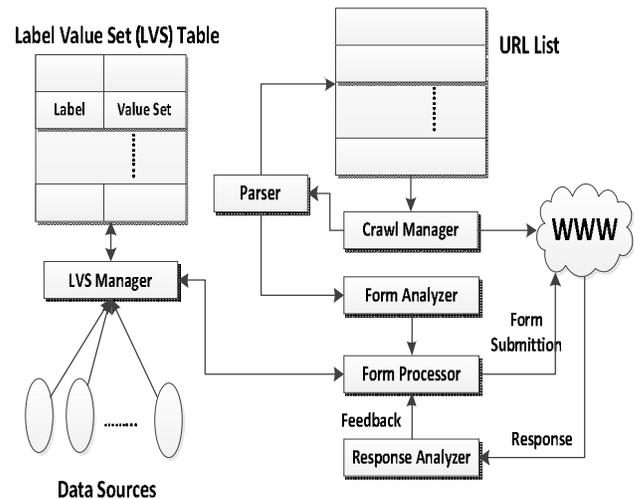

Fig. 5 Architecture of HiWE.

They modeled the form as having elements of the types: text box, select list, text area, radio button or checkbox. The domain of an element is the set of values that can be entered into this form element. In addition, each element is associated with a descriptive text termed as "label" for which it first finds the four closest texts to the element and then chooses one of them based on a set of heuristics defined by taking into accounting the relative position of each textual label. The candidate assignments for a form are generated from the values in the Label Value Set (LVS) table, which consists of (L, V) pairs, where "L" is a label and "V" is a fuzzy/grade set of the values belonging to this label. HiWE does not exhaust all of the possible assignments for a form. Although the authors have used the simple measure of the fraction of non-error pages returned, to evaluate each input, they assume the multiple-inputs to be independent and try to select specific URLs from the Cartesian product of inputs. Once a candidate assignment has been submitted to a form, HiWE caches the resulting page to the repository to support user queries. The major challenge of their approach is dealing with the form elements with infinite domain.

In 2002, Liddle et al. described a method to detect form elements and fabricate a HTTP GET request using default values specified for each field [9]. The proposed algorithm as depicted by the flowchart in Fig. 6 is not fully automated like HiWE and takes user input when required. Though the approach also models HTML forms in the same way as HiWE but is much simpler. The major contribution of the work is retrieving all or at least a significant percentage of the data before submitting all the queries. Without exhaustively trying all possible queries, the approach still extracts sufficient data. In the first phase of the approach, a default query is issued after which the site is sampled to determine if the response retrieved from the default query is comprehensive and finally the queries are issued exhaustively till the specified threshold is achieved. They suggested the use of stratified sampling method to select the candidate assignments that are most likely to extract new information from the hidden web site. The essence of the stratified sampling method is to use all form elements evenly.

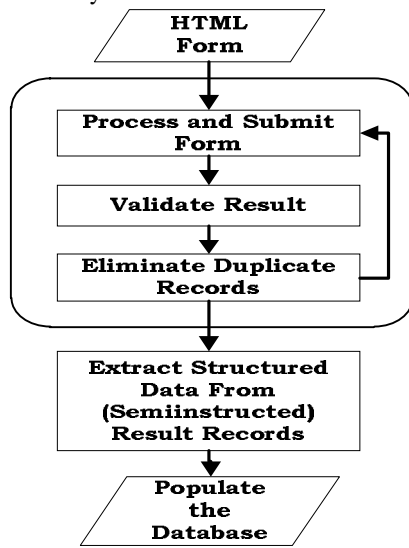

Fig. 6 Flowchart of liddele approach

Wu et al. in 2006 focus on the core issue of enabling efficient crawling of structured databases on the Web through iteratively issuing meaningful queries [4]. They proposed a theoretical framework that transforms the database crawling problem into a one of graph-traversal by following "relational" links. In their method, the structured web database DB is viewed as a single relational table with n data records {t1, t2,.....tn} over a set of m attributes { a1, a2, …am}. All distinct attribute values occurring in DB are contained by the Distinct Attribute Value set (DAV). Based on a data source DB, an attribute-value undirected graph (AVG) can be constructed. Each vertex vi є V represents a distinct attribute value avi є DAV and each undirected edge (vi; vj) stands for the coexistence of the two attribute values avi and avj in a record tk. According to AV G, the process of crawling is transformed into a graph traversal in which the crawler starts with a set of seed vertices and at each iteration a previously seen vertex v is selected to visit, thus all directly-connected new vertices and the records containing them are discovered and stored for future visits. However the attributes chosen in different queries can be different. It has been assumed that records and their different attributes can be extracted from the result pages to maintain reasonable coverage.

Madhavan et al. in 2009 discusses the approach used by Google in filling Web forms [5].HTML forms usually offer more than one input and hence a layman's strategy of enumerating the Cartesian product to identify of all possible inputs can result in a very large search space. They have presented an algorithm that appropriately chooses the input combinations so as to efficiently navigate the search space by including only those generated URLs which seem suitable for inclusion in the web search index. The first step of the approach contributes the in formativeness test for evaluating the query templates, i.e., combinations of form inputs. The basic idea of the in formativeness test is that all templates are probed to check which can return sufficiently distinct documents. The next step develops an algorithm that efficiently traverses the space of query templates to identify the ones suitable for surfacing. A template that returns enough distinct documents is deemed a good candidate for crawling. As a last step the approach contributes to an algorithm which predicts appropriate input values for the various form fields. They have described how the identification of typed inputs in web forms (e.g. zip codes, dates, prices) contributes to a better crawl.

In [7], a domain specific crawler for the hidden web, DSHWC that considers multi-input search forms has been developed. The working of DSHWC has been divided into several phases with the first one concerning the automatic downloading of the search forms. Phase 2 describes the most important component Domain-specific Interface Mapper that automatically identifies the semantic relationships between attributes of different search interfaces and guides the next step of merging the interfaces so as to form a Unified Search Interface (USI). The USI produced thereof is filled automatically and submitted to the Web. After obtaining response pages, the DSHWC stores the downloaded pages into Page repository that maintains the documents crawled/updated by the DSHWC along with their URLs. DSHWC [7] is a fully automated crawler which aims to obtain the response pages from Hidden Web by submitting filled search forms.

### B. Depth-Oriented Crawlers for Unstructured databases

Lot of research has been done in automating the retrieval of data hidden behind keyword based simple search forms which has been reviewed in this section.

Gravano et.al. in 2003 in their work in [14] presented a technique to automate the extraction of data from searchable text databases by taking a biased sample of documents that have been extracted by adaptively probing the database with topically focused queries. The queries have been automatically derived by using a classifier on a Yahoo! Like hierarchy of topics. The approach also evaluates the results and exploits the statistical properties of text thereof to derive frequency estimates for the words in extracted documents. The approach further suggested the use of focused probing for the classification of databases into a topic hierarchy. They have attempted to automatically categorize Hidden Web Databases by using a rule-based document classifier during probing.

In 2004, Barbosa and Freire in [6] claimed that assigning the values to fields of certain types like radio buttons, combo box is a bit easier than dealing with those that accept free form text as input like text boxes as these form elements actually expose the set of all possible values that can be input and automatically submitted by the crawler. They proposed a two phase algorithm to generate textual queries. The first stage involves creating a sample of data from the website and automatically selecting keywords which are associated weights based on the generated sample. This results in high recall; it then uses these keywords to build queries that siphon the results from the database in its second phase. To siphon, it uses a greedy algorithm so as to retrieve as much contents as possible with minimum number of queries, it iteratively selects the term with the highest frequency from the term list, and adds it to a disjunctive query if it leads to an increase in coverage. They have evaluated their algorithm over several real Web sites and obtained promising results in the preliminary stage itself. The results clearly indicated that their approach is effective in obtaining coverage of over 90% for most of the sites considered.

In 2005, Ntoulas et.al in their work [12] have provided a theoretical framework for analyzing the process of generating queries for a document collection that support single-attribute queries by examining the obtained results. The approach defines three policies for choosing the queries: a random policy where queries are randomly selected from a dictionary and serves as baseline for comparison, a generic policy based on the frequencies of keywords in any generic document corpus and an adaptive policy that learns from the collection downloaded so far. The process starts by learning a global picture starting with a random query, downloading the matched documents, and learning the next query from the current documents. This process is repeated until all the documents are downloaded.  They compared their adaptive method with two other query selection methods: the random method (queries are randomly selected from a dictionary), and the generic-frequency method (queries are selected from a 5.5-million-web-page corpus based on their decreasing frequencies). The experimental result shows that the adaptive method performs remarkably well in all cases.

Though much research focuses on the design and development of depth oriented hidden web crawlers but few have also focused on the issue of discovering relevant hidden web resources in a domain. This section presents a brief overview of some of the most cited works in this direction of hidden web crawlers.

### 1)  Breadth Oriented crawlers

In 2003, Bergholz et.al [10] focused on automatically discovering the entry points into the Hidden Web. They implemented a domain-specific crawling technique that starts out on the Surface Web using a general-purpose search engine to identify Hidden Web resources relevant in a domain. The crawling techniques to detect query able pages have been implemented and a method that help to assess whether a query able page is an HW resource or not has been developed. In their paper a Hidden Web crawler that discovers potentially interesting pages, analyzes and probes them to determine which pages can serve as Hidden Web resources has been described. Also, Experiments have been conducted to show that the number of Hidden Web resources is highly domain dependent, which can be found with little crawling effort. Their techniques perform well in both the domain-specific and random mode of crawling. The current crawler also combines the syntactic analysis of HTML forms with the query probing and show excellent results for full-text document search which comprises of a major portion of the Hidden Web but fails for a small fraction of the Hidden Web relevant to multi attribute form.

Barbosa and Freire in [11] in 2005 presented a form focused crawler (FFC) to automatically locate web forms based on topics. The architecture of FFC has been presented in Fig. 7. The crawler combines the use of a page classifier and a link classifier that have been trained for focusing its crawl on a particular topic by taking into account the contents of pages and patterns in & around the hyperlinks paths to a web page. The authors first make use of a backward search strategy to analyze and prioritize links which are likely to lead to a searchable form in one or more steps. The frontier manager is another major component of the FFC framework and is used to select the next target link for crawling based on their reward values decided by the current status of the crawler and the priority of the link in the current crawling step. The FFC also uses a form classifier to filter out useless forms. If a form is found searchable by the form classifier, it is added to the form database if not already present in it.

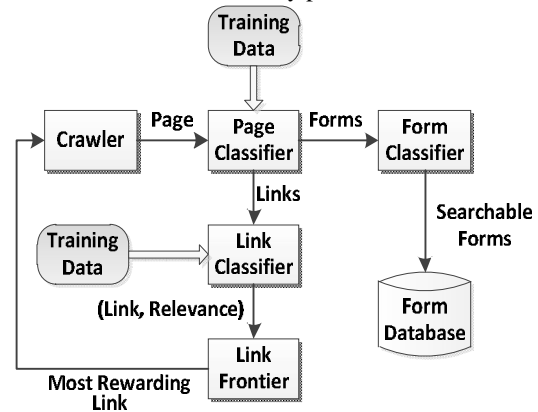

Fig. 7  Form Focused crawler

In 2006 Barbosa and Friere in [13] addressed the limitations of the FFC by presenting a new framework ACHE (Adaptive Crawler for Hidden-Web Entries) whereby crawlers adapt to their environments and improve the behavior by learning from previous experiences. Given a set of Web forms that are entry points to online databases, ACHE aims to efficiently and automatically locate other forms in the same domain.

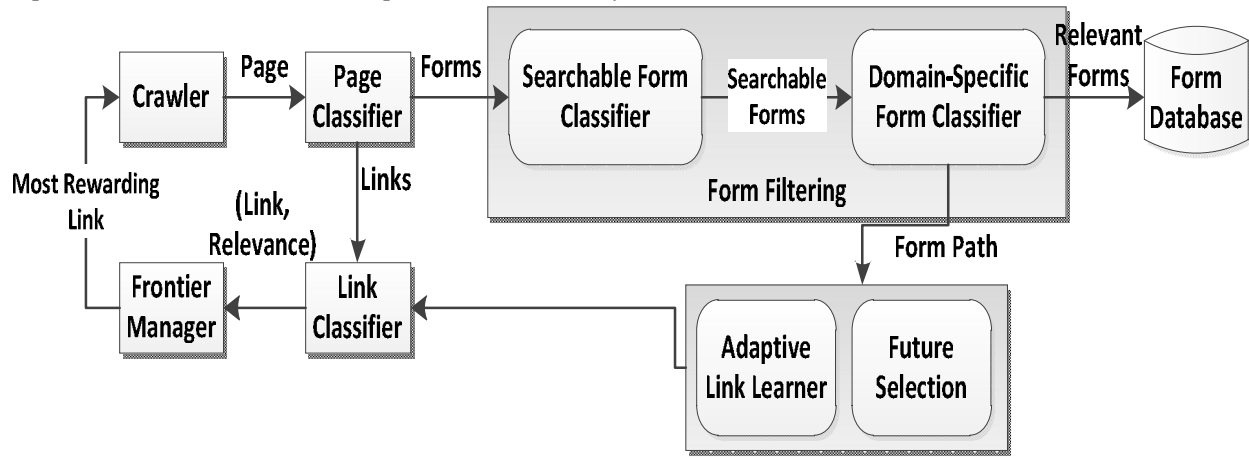

Fig. 8 ACHE Architecture

In addition to FFC, the ACHE comprises of two more classifiers: the searchable form classifier (SFC) which classifies the retrieved form as searchable or non-searchable and the domain-specific form classifier (DSFC) which checks whether the form belongs to the target domain. ACHE also employs a component called the adaptive link learner that dynamically learns features automatically extracted from successful paths by the feature selection component and updates the link classifier.

To achieve a notable progress in this fragment of Hidden Web crawling requires additional efforts for extending the current crawlers. In the next section we provide a comparison of the above discussed crawlers.

## VI. COMPARISON

Hidden Web crawlers are designed to automatically parse, process and interact with search forms. The tasks are automated by different crawlers in different ways which forms the focus of this study. A detailed comparison of the various Hidden Web crawlers that have been used in this study are been outlined in Table 1.

TABLE I
COMPARISON OF HIDDEN WEB CRAWLER

| Descriptive criteria | Year | Focused Perspective | Database type | Technique | Strength | Limitation |
|---|---|---|---|---|---|---|
| Raghavan et.al.[3] | 2001 | Depth-Oriented crawler for content extraction | Multi-attribute or structured | 1) Text similarity to match fields and domain attributes. 2) Partial page layout and visual adjacency for identifying form elements 3) Hash of visually important parts of the page to detect errors | 1) Significant contribution to label matching process 2) Updates the user provided task description by learning information from the successful extracts of crawling. | 1) ignores forms with fewer than 3 attributes 2) Require significant human input thus performance highly depends on the quality of input data       3) not scalable to hidden web databases in diversified domains. |
| Liddle et.al. [9] | 2002 | Depth-Oriented crawler for content extraction | Multi-attribute or structured | 1) Stratified Sampling Method (avoid queries biased toward certain fields) 2)Fields with finite set of values, ignores automatic filling of text field 3) Concatenation of pages connected through navigational elements | 1) domain-independent approach 2) accounts for duplicate results identified by computing hash values | 1) Do not consider detection of forms inside result pages. 2) Detection of record boundaries and computes hash values for each sentence poses huge resource requirements. |
| Garvano et.al. [14] | 2002 | Depth-Oriented crawler for content extraction | document based or unstructured | 1) use of topically focused queries 2) adaptive query probing | 1) facilitates design of meta-search engines 2)  used to categorize hidden web databases | 1) Query chosen only by using hierarchical categories as in Yahoo! and does not consider flat classification |
| Bergholz et.al. [10] | 2003 | Breadth-oriented | unstructured | 1) domain specific crawling 2) Query prober to recognize and | 1) Efficient at discovering unstructured hidden web resources as uses the | 1) Only deal with full text search forms. 2) Initialized with pre-classified |

| | | | | | |
|---|---|---|---|---|---|
| | crawler for resource discovery | databases in a domain | assess the usefulness of the HW resource. | combination of syntactic elements of HTML forms and query probing technique. | documents and relevant keywords |
| Barbosaet.al. [6] | 2004 | Depth-Oriented crawler for content extraction | document based or unstructured | 1) Considers candidate query based on its frequency of appearance in each round | 1) Simple and completely automated strategy 2) Automatically creates sufficiently accurate description of document therefore, can be used in other resource discovery systems. 3) Leads to high coverage. | 1) No assurance of acquiring new pages 2) ineffective for search interfaces that fix the number of returned results 3) simple approach therefore raises security issues |
| Ntoulas et.al. [12] | 2005 | Depth-Oriented crawler for content extraction | document based or unstructured | 1) Incremental adaptive method 2) frequency estimation based on already downloaded documents 3) greedy algorithm that tries to maximize the 'potential gain' in every step. | 1) Combination of policies (random, generic and adaptive) for choosing appropriate queries. 2) use of multiple frequency estimators -independent and zipf's law based | 1) Query distribution does not make sure to adapt to the attribute values set of the database. 2) Memory requirements for calculating potential gain are huge. 3) Assumed constant cost for every query which does not hold in real situations. |
| Barbosa et.al. [11] | 2005 | Breadth-oriented crawler for resource discovery | structured & unstructured databases | 1) Link classifier to focus search on a specific topic 2) use of a stopping criteria to avoid unproductive searches | 1) Highly efficient in retrieving searchable forms focused for a particular topic | 1) Manually selecting a representative training set is difficult so creating the link classifier is time consuming |
| Alvarez et.al. [16] | 2006 | | | 1) set of domain definitions each one of which describes a data-collection task 2) use of heuristics to automatically identify relevant query forms | 1) System can be extended for discovering relevant resources. 2) Handles client side as well as server side hidden Web 3) Experimentally proved effective for collecting data. | 1) No defined threshold for associating form elements and attributes in the domain definitions 2) hypothetical assumption of having at least one label associated with every form element which does not hold true for most of the bounded form elements (drop down boxes) |
| Ping Wu et.al. [4] | 2006 | Depth-Oriented crawler for content extraction | Multi-attribute or structured | 1) Models each structured database as a distinct attribute -value graph 2) Set the graph to crawl the database (set-covering problem) | 1) issues only meaningful queries as tuned with domain knowledge 2) overcomes limitation of greedy methods | 1) Query results in each round must be added to the graph thus involves huge cost of resources |
| Barbosa et.al. [13] | 2007 | Breadth-oriented crawler for resource discovery | unstructured databases | 1)Greedy algo derived by the weights associated to keywords in the collected data 2)Issue queries using dummy words to detect error pages | 1) Improved harvest rates as crawl progresses 2) retrieves homogeneous set of forms 3) Automated and adaptive thus eliminates any bias arising out of learning process. | 1) configuring the crawler to start initially needs more effort than manually configured crawlers 2) works only for Single keyword-based queries |
| Madhavan et.al. [5] | 2008 | Depth-Oriented crawler for content extraction | Multi-attribute or structured | 1) Evaluate the query templates by defining the in formativeness test. | 1) efficiently navigates the search space of possible input combinations | 1) No consideration to the efficiency of deep web crawling |
| Komal Bhatia et.al. [7] | 2010 | Depth-Oriented crawler for content extraction | Multi-attribute or structured | 1) Domain Specific Interface Mapper to create unified query interfaces for a domain 2) calculation of re-visit frequency based on probability of change of web page | 1) Multi-strategy interface matching 2) use of mapping knowledge base to avoid repetition for minimizing the mapping effort 3) Enhances the scope of developing a specialized search engine for the Hidden Web. | 1) Indexing technique was not specified for storing pages in the repository 2) Defined the performance only for crawling while the efficiency of schema matching and merging procedures over variety of query interfaces has not been quantified. |
| Sonali Gupta et.al [8] | 2013 | Depth-Oriented crawler for content extraction | document based or unstructured | 1) Creates a domain representation that is stored in domain specific data repository. 2) uses a domain specific classification hierarchy for query term identification | 1) Achieves high coverage with just a small number of queries 2) makes use of domain specific data repositories and thus can be extended to other domains 3) Can be fully automated if integrated with semantic web technologies. | 1) Requires human effort for an initial start of the crawler. 2) domain-specific |

## VII. CONCLUSIONS

Hidden Web crawlers enable indexing, analysis and mining of hidden web content. The extracted content can then be used to categorize and classify the hidden databases. The paper discusses the various crawlers that have been developed for surfacing the contents in the Hidden Web. The crawlers have also been differentiated on the basis of their underlying techniques and behavior towards different kind of search forms and domains. As each of the discussed crawlers have their own strengths and limitations, much more needs to be explored in the area for better research prospective.

## REFERENCES

[1] Michael Bergman, "The deep Web: surfacing hidden value". In the Journal Of Electronic Publishing 7(1) (2001).

[2] Sonali Gupta, Komal Kumar Bhatia: Exploring 'Hidden' parts of the Web: the Hidden Web, in 4rth International Conference on Advances in recent technologies in communication and computing, ARTCom 2012 proceedings in Lecture Notes in Electrical Engineering , Springer Verlag Berlin Heidelberg , ISSN 1876-1100, p.p. 508-515, 2012.

[3] S. Raghavan, H. Garcia-Molina. Crawling the Hidden Web. In: the proceedings of the 27th International Conference on Very large databases VLDB'01, Morgan Kaufmann Publishers Inc., San Francisco, CA, p.p. 129-138.

[4] Ping Wu, J.-R. Wen, H. Liu, and W.-Y. Ma. Query Selection Techniques for Efficient Crawling of Structured Web Sources. In ICDE, 2006

[5] J. Madhavan, D. Ko, L. Kot, V. Ganapathy, A. Rasmussen, A. Halevy : Google's Deep-Web Crawl. In proceedings of Very large data bases VLDB endowment, pp. 1241-1252, Aug. 2008.

[6] L. Barbosa, J. Freire : Siphoning hidden-web data through keyword-based interfaces. In: SBBD, 2004, Brasilia, Brazil, pp. 309-321.

[7] Komal kumar Bhatia, A.K.Shrma, Rosy Madaan: AKSHR: A Novel Framework for a Domain-specfic Hidden web crawler. In Proceedings of the first international Conference on Parallel, Distributed and Grid Computing, 2010.

[8] Sonali Gupta, Komal Kumar Bhatia: HiCrawl: A Hidden Web crawler for Medical Domain in proceedings of 2013 IEEE International Symposium on Computing and Business Intelligence, ISCBI, August18-18, 2013 Delhi , India .

[9] S. W. Liddle, D. W. Embley, D. T. Scott, S. H. Yau. Extracting Data Behind Web Forms. In: 28th VLDB Conference2002 , HongKong, China.

[10] A. Bergholz, B. Chidlovskii. Crawling for domain-specific Hidden Web resources. In Proceedings of the Fourth International Conference on Web Information Systems Engineering (WISE'03). pp.125-133 IEEE Press, 2003

[11] L. Barbosa and J. Freire. Searching for Hidden-Web Databases. In Proceedings of WebDB, pages 1–6, 2005.

[12] A. Ntoulas, P. Zerfos, J.Cho. Downloading Textual Hidden Web Content Through Keyword Queries. In: 5th ACM/IEEE Joint Conference on Digital Libraries (Denver, USA, Jun 2005) JCDL05, pp. 100-109.

[13] L.Barbosa and J.Freire, An adaptive crawler for locating hidden-web entry points," in Proc. of WWW, 2007, pp. 441-450.

[14] P.Ipeirotis and L. Gravano, Distributed search over the hidden web: Hierarchical database sampling and selection," in VLDB, 2002.

[15] K.C. Chang, B. He, M.Patel, Z.Zhang : Structured Databases on the Web: Observations and Implications. SIGMOD Record, 33(3). 2004.

[16] B. He, M.Patel, Z.Zhang, K.C. Chang: Accessing the Deep Web: A survey. Communications

[17] of the ACM, 50(5):95–101, 2007.

[18] Manuel Álvarez, Juan Raposo, Alberto Pan, Fidel Cacheda, Fernando Bellas, Víctor Carneiro: Crawling the Content Hidden Behind Web Forms. In Proceedings of the 2007 International conference on Computational Science and its applications, Published by Springer-Verlag Berlin, Heidelberg, 2007.